\documentclass[a4paper]{article}

\usepackage{amssymb}
\usepackage{amsfonts}
\usepackage{amsmath}
\usepackage{graphicx}
\usepackage{hyperref}
\hypersetup{backref,
            pagebackref=true,
            ps2pdf,
            bookmarks=true,
            bookmarksnumbered=true,
            pdfauthor=,
            colorlinks=true,
            linkcolor=blue}

\title{Coulomb-Gas Approach for Percolation Theory }
\author{Smain BALASKA \thanks{%
sbalaska@yahoo.com ,balaska.smain@univ-oran.dz } and \textit{\ }Toufik SAHABI \thanks{%
sahabitoufik@yahoo.fr\ \ } \\
\textit{Laboratoire de Physique Th\'{e}orique d'Oran. }\\
\textit{Dpt de Physique. Universit\'{e} d'Oran.}\\
\textit{BP 1524 El'Manouer 31000 Oran. Alg\'{e}ria.}}

\begin{document}
\maketitle

\begin{abstract}
The aim of this work is to present a non trivial confirmation of
the powerful of the Coulomb gas-techniques for Boundary Conformal
Field Theory (BCFT). We show that we can re-derive the known Cardy
result of percolation problem via the technics developed by S.
Kawai in the Coulomb-Gas formalism.

PACS numbers : 11.25.Hf
\end{abstract}

\section{Critical percolation and conformal invariance}
On BCFTs, Coulomb gas formalism presents a strong tool to obtain
correlation functions without having to solve differential
equations not only in the case of conformal minimal models but
also in different models such as percolation model. In his papers
\cite{shin1}\cite{shin2}, S. Kawai presents a general formalism to
compute correlation functions in the half plane using the
free-field construction of boundary states and applying the
Coulomb gas formalism. This formalism was applied for the critical
Ising model with free and fixed boundary conditions obtained from
Cardy's boundary states. In this work, we will use Kawai's
techniques to provide the percolation crossing probability which
is a two-point correlation function on the upper half plane.

\begin{figure}
\centering
    \includegraphics[scale=0.5]{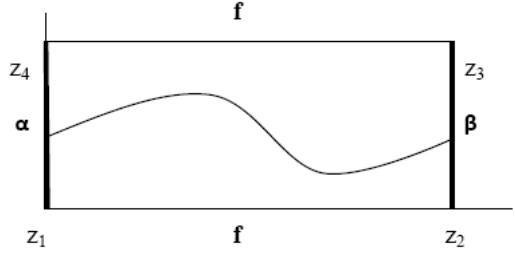}
    \caption{Crossing cluster between the two segments of  the rectangle}
    \label{figigure}
\end{figure}

In the thermodynamic limit, and of course, at the critical point,
percolation is believed to be described by a conformal field theory $%
M(p^{\prime }=2,p=3)$ with vanishing central charge \cite{cardy1}.
Crossing probability is of great interest in studies of
percolation. In two dimensions and in geometries with edges (a
rectangle for example), a crossing event is a configuration of
bonds or sites on the lattice covering this geometry such that
there exists at least one cluster connecting two disjoint segments
of the boundary. See figure \ref{figigure} for the case of the
rectangle.

Let $\pi $ be the crossing probability associated with this event,
and $p$ be the probability for each bond to be open (and $1-p$ to
be closed).Then there is a critical value of $p$\ called the
percolation threshold such as
\begin{equation}
\pi =\left\{
\begin{array}{c}
0\text{ \ \ \ \ \ \ \ \ \ \ \ \ if \ \ \ }p<p_{c} \\
\pi (p,r)\text{ \ \ \ if \ \ }p=p_{c} \\
1\text{ \ \ \ \ \ \ \ \ \ \ \ \ if \ \ \ }p>p_{c}%
\end{array}%
\right\}
\end{equation}%
where $r$ represents the aspect ratio of the rectangle
(height/width). The most familiar way to think about percolation
as a critical phenomenon is
through the $Q\rightarrow 1$ limit of the $Q$-state Potts model. Let $%
Z_{\alpha \beta }$\ be the partition function of the $Q$-state Potts model
with the constraint that all spins at lattice sites on $S_{1}$ are fixed in
the state $\alpha $ and all the spins on $S_{2}$ are fixed in the state $%
\beta $ . The boundary spins are free on the horizontal sides of the
rectangle. The existence of a cluster would force spins on the two segments (%
$S_{1},S_{2}$) to be in the same state. In this way,
a formula for $\pi (p,r)$ on the lattice is obtained as \cite{cardy2},\cite%
{langlands}%
\begin{equation}
\pi (p,r)=\lim_{Q\rightarrow 1}(Z_{\alpha \alpha }-Z_{\alpha \beta })
\label{eq2}
\end{equation}%
where the partition functions we need are given in terms of correlators by%
\begin{eqnarray}
Z_{\alpha \alpha } &=&Z_{f}\left\langle \phi _{(f\alpha )}(z_{1})\phi
_{(\alpha f)}(z_{2})\phi _{(f\alpha )}(z_{3})\phi _{(\alpha
f)}(z_{4})\right\rangle   \nonumber \\
Z_{\alpha \beta } &=&Z_{f}\left\langle \phi _{(f\alpha )}(z_{1})\phi
_{(\alpha f)}(z_{2})\phi _{(f\beta )}(z_{3})\phi _{(\beta
f)}(z_{4})\right\rangle   \label{eq3}
\end{eqnarray}%
where $Z_{f}$ is the partition function with free boundary conditions and $%
\phi _{(ij)}$ denote the boundary operator corresponding to a switch from
boundary condition $(i)$ to $(j)$ at the point $x.$In our case the relevant
boundary changing operator is identified as being the $\phi _{(12)}$
boundary primary field in the $M(2,3)$ theory ($Q\rightarrow 1$ limit) \cite%
{Mathieu}, \cite{rasmussen}. For this particular minimal model we see, from
the Kac table formula%
\begin{eqnarray}
h_{r,s} &=&\frac{[r(m+1)-sm]^{2}-1}{4m(m+1)}  \nonumber \\
r,s &>&0\text{ \ \ \ \ \ , \ \ \ \ \ }m=2  \label{eq4}
\end{eqnarray}%
that $h_{1,2}=0$ and we have in addition $\lim_{Q\rightarrow 1}Z_{f}=1$.
Then, to obtain the crossing probability formula one has to find the form of
the four-point correlation functions \cite{Watts}%
\begin{equation}
G(\eta )=\left\langle \phi _{(12)}(z_{1})\phi _{(12)}(z_{2})\phi
_{(12)}(z_{3})\phi _{(12)}(z_{4})\right\rangle   \label{eq5}
\end{equation}%
which will depend only on the cross ratio
\[
\eta =\frac{(z_{1}-z_{2})(z_{3}-z_{4})}{(z_{1}-z_{3})(z_{2}-z_{4})}
\]

\section{Correlation functions from the Coulomb-Gas approach}

The correlator (\ref{eq5}) can be written in the upper half plane- by
application of the method of images - (see \cite{Difrancesco}) as :%
\begin{eqnarray}
G(\eta ) &=&\left\langle \phi _{(12)}(z_{1})\phi _{(12)}(z_{2})\phi
_{(12)}(z_{3})\phi _{(12)}(z_{4})\right\rangle  \nonumber \\
&=&\left\langle \phi _{(12)}(z_{1},\overline{z}_{1})\phi _{(12)}(z_{2},%
\overline{z}_{2})\right\rangle _{UHP}  \label{eq6}
\end{eqnarray}%
where $z_{3}=\overline{z}_{1}$ and $z_{4}=\overline{z}_{2}.$

In the Coulomb gas approach the boundary 2-point correlation function for
physical boundary conditions can be obtained by introducing the screened
vertex operators \cite{shin1}, \cite{shin2}. In this case the correlator (%
\ref{eq6}) becomes
\[
G(\eta )=\left\langle B(\alpha )\left\vert
V_{(12)}^{m_{1},n_{1}}(z_{1})V_{(12)}^{\overline{m}_{1},\overline{n}_{1}}(%
\overline{z}_{1})V_{(12)}^{m_{2},n_{2}}(z_{2})V_{(12)}^{\overline{m}_{2},%
\overline{n}_{2}}(\overline{z}_{2})\right\vert 0,0;\alpha _{0}\right\rangle
\]%
where $\left\vert B(\alpha )\right\rangle $ is called the boundary coherent
state, $\left\vert 0,0;\alpha _{0}\right\rangle $ represents the vacum state
and $V$ and $\overline{V}$ are the screened Vertex operators.

In order that the correlator be non-vanishing, we must satisfy the charge
neutrality conditions
\begin{equation}
m+\overline{m}=0\text{ \ \ \ \ \ \ , \ \ \ \ \ \ \ \ }n+\overline{n}=1
\label{eq7}
\end{equation}

then we have the boundary charge $\alpha =\alpha _{1,1}=0$ for the first
condition and $\alpha =\alpha _{1,3}=-\alpha _{-}$ for the second one, where%
\begin{eqnarray*}
\alpha _{r,s} &=&\frac{1}{2}(1-r)\alpha _{+}+\frac{1}{2}(1-s)\alpha _{-} \\
\alpha _{+} &=&\sqrt{\frac{p}{p^{\prime }}}=\sqrt{\frac{3}{2}} \\
\alpha _{-} &=&-\sqrt{\frac{p^{\prime }}{p}}=-\sqrt{\frac{2}{3}}
\end{eqnarray*}

The first condition (\ref{eq7}) corresponds to the conformal block%
\begin{eqnarray}
I_{1} &=&N_{1}\left\{ \frac{(z_{1}-\overline{z}_{1})(\overline{z}_{2}-z_{2})%
}{(z_{1}-z_{2})(\overline{z}_{1}-\overline{z}_{2})(z_{1}-\overline{z}_{2})(%
\overline{z}_{1}-z_{2})}\right\} ^{2h_{1,2}}\frac{\Gamma (1-\alpha
_{-}^{2})^{2}}{\Gamma (2-2\alpha _{-}^{2})}  \nonumber \\
&&\text{ \ \ \ \ \ \ }\times F(2b,1-\alpha _{-}^{2},2-\alpha _{-}^{2};\eta )
\nonumber \\
&=&N_{1}\frac{\Gamma (\frac{1}{3})^{2}}{\Gamma (\frac{2}{3})}F(0,\frac{1}{3},%
\frac{4}{3};\eta )  \label{eq8}
\end{eqnarray}%
where $N_{1}$ is a constant, $b=2\alpha _{1,2}(2\alpha _{0}-\alpha _{1,2})$
with $\alpha _{0}=\frac{1}{2\sqrt{6}}$ and $F$ is the hypergeometric
function of the Gaussian type. Whereas the second condition corresponds to
the conformal block%
\begin{eqnarray}
I_{2} &=&N_{2}\left\{ \frac{(z_{1}-\overline{z}_{1})(\overline{z}_{2}-z_{2})%
}{(z_{1}-z_{2})(\overline{z}_{1}-\overline{z}_{2})(z_{1}-\overline{z}_{2})(%
\overline{z}_{1}-z_{2})}\right\} ^{2h_{1,2}}\frac{\Gamma (1-\alpha
_{-}^{2})\Gamma (3\alpha _{-}^{2}-1)}{\Gamma (2\alpha _{-}^{2})}  \nonumber
\\
&&\text{ \ \ \ \ \ \ \ \ }\times (-\eta )^{2h_{1,2}+\frac{\alpha _{-}^{2}}{2}%
}\text{ }F(\alpha _{-}^{2},1-\alpha _{-}^{2},2\alpha _{-}^{2};\eta )
\nonumber \\
&=&N_{2}\frac{\Gamma (\frac{1}{3})\Gamma (1)}{\Gamma (\frac{4}{3})}(-\eta )^{%
\frac{1}{3}}\text{ }F(\frac{2}{3},\frac{1}{3},\frac{4}{3};\eta )  \label{eq9}
\end{eqnarray}%
where $N_{2}$ is a constant. Using the properties of the $\Gamma $ and the
hypergeometric functions \cite{gradshteyn} we can write the conformal blocks
of equations (\ref{eq8}) and (\ref{eq9}) as
\begin{equation}
I_{1}=N_{1}\frac{\Gamma (\frac{1}{3})^{2}}{\Gamma (\frac{2}{3})}
\label{eq10}
\end{equation}

\begin{equation}
I_{2}=-3N_{2}(\eta )^{\frac{1}{3}}\text{ }F(\frac{1}{3},\frac{2}{3},\frac{4}{%
3};\eta )  \label{eq11}
\end{equation}

To find the appropriate combination (i.e the values of $N_{1}$ and $N_{2}$)
describing $\pi (p,r),$we give the precise correspondence between the aspect
ratio $r$ of the rectangle and the cross ratio $\eta $ by the two expressions%
\begin{equation}
r=\frac{-iz_{4}}{z_{2}}\text{ \ \ \ \ \ \ and \ \ \ \ \ \ }\eta =\frac{%
z_{2}^{2}}{\left\vert z_{3}\right\vert ^{2}}  \label{eq12}
\end{equation}

For infinitly wide lattice ($r\rightarrow 0$ and $\eta \rightarrow 1$), the
vertical crossing probability $\pi _{v}(p,r)$ should be 1. But for infinitly
narrow lattice ($r\rightarrow \infty $ and $\eta \rightarrow 0$) it should
be zero. Thus we find $N_{1}=0$ and $N_{2}=-\Gamma (\frac{2}{3})/\Gamma (%
\frac{1}{3})^{2}.$ Then the vertical crossing probability take the form
\begin{equation}
\pi _{v}(p,r)=\pi ((z_{1},z_{2});(z_{3},z_{4}))=3\frac{\Gamma (\frac{2}{3})}{%
\Gamma (\frac{1}{3})^{2}}(\eta )^{\frac{1}{3}}\text{ }F(\frac{1}{3},\frac{2}{%
3},\frac{4}{3};\eta )  \label{eq13}
\end{equation}

The horizontal crossing probability $\pi _{h}(p,r)$ should be 1 for
infinitly narrow lattice and zero for infinitly wide lattice. Then we have
in this case $N_{1}=N_{2}=\Gamma (\frac{2}{3})/\Gamma (\frac{1}{3})^{2}$ and
we can write%
\begin{equation}
\pi _{h}(p,r)=\pi ((z_{1},z_{4});(z_{2},z_{3}))=1-3\frac{\Gamma (\frac{2}{3})%
}{\Gamma (\frac{1}{3})^{2}}(\eta )^{\frac{1}{3}}\text{ }F(\frac{1}{3},\frac{2%
}{3},\frac{4}{3};\eta )  \label{eq14}
\end{equation}

Of course, $\pi _{h}+\pi _{v}=1$ which means that whenever there is an
horizontal cluster, it cannot exist a vertical one. The two events are
incompatible.

If we change the labelling of the rectangle corner's from ($%
z_{1},z_{2},z_{3},z_{4}$ ) to ($z_{2},z_{3},z_{4},z_{1}$) , we retreview
Cardy's results \cite{cardy1}

\begin{equation}
\pi _{v}(p,r)=\pi ((z_{1},z_{4});(z_{2},z_{3}))=1-3\frac{\Gamma (\frac{2}{3})%
}{\Gamma (\frac{1}{3})^{2}}(1-\eta )^{\frac{1}{3}}\text{ }F(\frac{1}{3},%
\frac{2}{3},\frac{4}{3};1-\eta )  \label{eq16}
\end{equation}%
and

\begin{equation}
\pi _{h}(p,r)=\pi ((z_{1},z_{2});(z_{3},z_{4}))=3\frac{\Gamma (\frac{2}{3})}{%
\Gamma (\frac{1}{3})^{2}}(1-\eta )^{\frac{1}{3}}\text{ }F(\frac{1}{3},\frac{2%
}{3},\frac{4}{3};1-\eta )  \label{eq17}
\end{equation}

and from equations (\ref{eq13}), (\ref{eq14}), (\ref{eq16}) and (\ref{eq17})
we retreview also that
\begin{equation}
\pi (\eta )=1-\pi (1-\eta )
\end{equation}

\textbf{Acknowledgments\newline
}

T. Sahabi would like to thank G. Watts for discussions on aspects of
percolation and crossing formulas and his comments on this work.

\end{document}